\documentclass[apjl]{emulateapj}
\usepackage{apjfonts}

\newcommand{\Msun}{\,\ensuremath{\mbox{M}_{\odot}}}

\newcommand{\vLSR}{\ensuremath{v_{\mbox{\scriptsize LSR}}}}
\newcommand{\kms}{\,\ensuremath{\rm{km\,s}^{-1}}}
\newcommand{\cm}{\,\ensuremath{\mbox{cm}^{-2}}}
\newcommand{\kpc}{\,\ensuremath{\mbox{kpc}}}
\newcommand{\mA}{\,\ensuremath{\mbox{m\AA}}}

\newcommand{\Ca}{\ensuremath{\mbox{\ion{Ca}{2}}}}
\newcommand{\CaK}{\ensuremath{\mbox{\ion{Ca}{2}~K}}}
\newcommand{\CaH}{\ensuremath{\mbox{\ion{Ca}{2}~H}}}
\newcommand{\CaHK}{\ensuremath{\mbox{\ion{Ca}{2}~H and K}}}
\newcommand{\NaI}{\ensuremath{\mbox{\ion{Na}{1}}}}
\newcommand{\NaD}{\ensuremath{\mbox{\ion{Na}{1}~D}}}
\newcommand{\NaDa}{\ensuremath{\mbox{\ion{Na}{1}~D$_{1}$}}}
\newcommand{\NaDb}{\ensuremath{\mbox{\ion{Na}{1}~D$_{2}$}}}

\newcommand{\HI}{\ensuremath{\mbox{\ion{H}{1}}}}

\newcommand{\teeWCaK}{W\ensuremath{_\lambda(\CaK) = 114.6\pm 4.4 \mA}}
\newcommand{\teeNHI}{N(\HI) = \ensuremath{1.64\pm0.31\times10^{19}\cm}}
\newcommand{\teeNCa}{N(\Ca) = \ensuremath{1.32\pm0.05\times10^{12}\cm}}

\newcommand{\teeNCaHI}{N(\Ca)/N(\HI) = \ensuremath{81 \pm 16 \times 10^{-9}}}
\newcommand{\heeWNaDa}{W\ensuremath{_\lambda(\NaDa) = 9.3\pm 4.4 \mA}}
\newcommand{\heeWNaDb}{W\ensuremath{_\lambda(\NaDb) = 12.1\pm 5.0 \mA}}

\newcommand{\tee}{\mbox{HE\,$1048+0231$}}
\newcommand{\hee}{\mbox{HE\,$1138-1303$}}
\newcommand{\wb}{Complex~WB}
\newcommand{\ubv}{\mbox{$U\!BV$}}

\shorttitle{The Distance to HVC Complex WB}
\shortauthors{Thom et al.}

\begin{document}

\title{The Galactic Nature of High Velocity Cloud Complex WB}


\author{C. Thom\altaffilmark{1,2}, M. E. Putman\altaffilmark{3},
B. K. Gibson\altaffilmark{1,4}, N. Christlieb\altaffilmark{5},
C. Flynn\altaffilmark{2}, T. C. Beers\altaffilmark{6},
R. Wilhelm\altaffilmark{7} {\scriptsize AND} Y.S. Lee\altaffilmark{6}}
\email{cthom@astro.swin.edu.au}

\altaffiltext{1}{Centre for Astrophysics and Supercomputing, Swinburne University of
  Technology, PO Box 218, Hawthorn, Victoria, 3122, Australia}

\altaffiltext{2}{Tuorla Observatory, V\"ais\"al\"antie 20, FI-21500, Piikki\"o,
Finland}

\altaffiltext{3}{Dept. of Astronomy, University of Michigan, 500 Church St., Ann Arbor, MI 48109}

\altaffiltext{4}{Centre for Astrophysics, University of Central Lancashire, Preston, PR1 2HE, UK}

\altaffiltext{5}{Hamburger Sternwarte, Universit\"at Hamburg, Gojenbergsweg 112,
  D-21029 Hamburg,Germany}

\altaffiltext{6}{Department of Physics \& Astronomy, and JINA: Joint Institute for Nuclear
Astrophysics, Michigan State University, E. Lansing, MI, 48824}

\altaffiltext{7}{Department of Physics, Texas Tech University, Lubbock, TX 79409}


\begin{abstract}

We have detected absorption lines from the High Velocity Cloud \wb\ in the
spectrum of the star \tee.  This detection sets an upper distance limit to
the cloud of $8.8^{+2.3}_{-1.3}\kpc$.  Non-detection (at $>4\sigma$ confidence) in
the star \hee\ at $7.7\pm 0.2\kpc$ sets a probable lower limit. The
equivalent width of the \CaK\ line due to the HVC (\teeWCaK) corresponds to
a column density of \teeNCa.  Using an \HI\ spectrum from the
Leiden/Argentine/Bonn survey, we calculated \teeNCaHI. These distance
limits imply an \HI\ mass limit of $3.8 \times 10^5\Msun < \mbox{M}_{\HI} <
4.9 \times 10^5\Msun$.  The upper distance limit imposed by these
observations shows that this HVC complex has a probable Galactic or
circum-Galactic origin. Future metallicity measurements will be able to
confirm or refute this interpretation.

\end{abstract}


\section{Introduction}

High Velocity Clouds (HVCs) are clouds of neutral hydrogen (\HI) gas with
velocities that are inconsistent with a simple model of differential
Galactic rotation. Since a kinematic distance cannot be derived, the exact
location and nature of the HVCs has been the topic of much
speculation. Determination of the distances to HVCs is vitally important,
as such measurements can constrain their likely origins, and establish
their physical parameters, many of which scale with distance.

The HVCs have been ascribed to many different origins, from local Galactic
processes, to distant proto-galaxies. \citet{shapiro-field-76} were the
first to suggest the existence of a Galactic Fountain, in which supernovae
drive hot gas into the halo, which then condenses and returns to the disk
\citep[e.g.][]{bregman-80}. Extra-Galactic origins have been discussed
almost since the discovery of HVCs \citep{oort-66-HVC-origins}.
\citet{blitz-etal-99} have inspired vigorous debate by reviving and further
refining this scenario, noting that it naturally explains the kinematics of
clouds towards the bary- and anti-barycenter of the Local Group. They
suggested that these distant HVCs could be dark-matter dominated objects,
which some have claimed might be the solution to the ``satellite problem''
of Cold Dark Matter simulations \citep[namely, that many more dark matter
haloes are predicted to exist than satellite galaxies observed,
e.g. ][]{klypin-etal-99-missing-satellites}.

Despite many years of effort, few HVCs have been detected in absorption
against stellar probes.  Of the traditional complexes, only M and A have
been detected
\citep{danly-etal-93-ComplexM,vanwoerden-etal-99-nature}. Several other
smaller clouds have also been detected in the spectra of background stars
\citep[e.g.][]{bates-etal-90-HVC-dist,sembach-savage-massa-91-HVC-dist}.
\citet{wakker-01-distance-metallicity} provides a much more comprehensive
summary.

\citet{wannier-wrixon-wilson-72} were the first to report on the positive
velocity clouds in the third and fourth Galactic quadrants.  Fainter than
the traditional negative velocity complexes such as A and C, they were
divided into four separate complexes WA -- WD \citep[][hereafter
WvW91]{wvw91}. These complexes span a wide range of position and velocity
space, comprising a total of 64 clouds in the WvW91 catalog, with mostly
moderate deviation velocities\footnote{Deviation velocity is the amount by
which a velocity deviates from allowed Galactic rotation velocities
\citep{wakker-91-II-distribution}.}. \wb\ ($l = 225 - 265\degr$; $b = 0 -
60\degr$) is composed of 29 separate clouds, with a total \HI\ mass
$\mbox{M(\HI)} = 0.236\,d^{2}(\kpc)\,S({\rm Jy}\kms) \sim 6.5\times
10^{3}\,d^{2}\Msun$, where $S$ is the integrated flux and $d$ is the
distance. \citet{robertson-etal-91-complexWB} have observed the Wannier
Clouds in absorption towards the Seyfert galaxy PKS\,0837--12. Along this
line of sight they derive a N(\Ca)/N(\HI) ratio of $160 \times 10^{-9}$.
Using a higher resolution \HI\ spectrum,
\citet{wakker-01-distance-metallicity} revised this value upwards to $280
\times 10^{-9}$. Both sets of authors note the anomalously high calcium
abundance (one of the highest measured for any HVC), while
\citeauthor{robertson-etal-91-complexWB} concluded that a distance is
necessary to discern between the various possible explanations they
offered.

The Wannier clouds, along with the extreme positive velocity clouds
(population EP) lie in the direction of the Local Group anti-barycenter.
They are explicitly included in the \citet{blitz-etal-99} scenario as being
very distant and undergoing infall towards the Local Group barycenter.
Thus a confirmed upper limit on the distance would stand in opposition to
this suggestion and may have broader consequences for the HVCs as a whole.

\section{Observations and Data}
\label{sec: observations}


\begin{figure}
  \epsscale{1.3} 
  \plotone{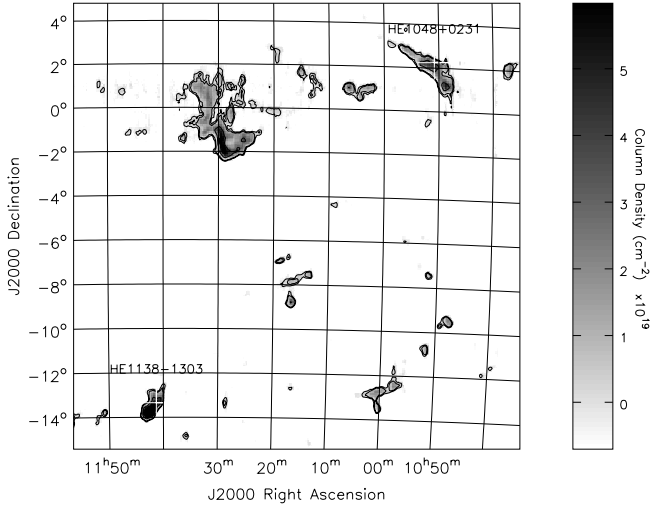}

  \caption{HIPASS column density map showing region around both sight lines
  (not the whole of \wb). The labelled white plus signs indicate the
  positions of the stars. Contours are drawn at $8.0 \times 10^{18}, 1.0
  \times 10^{19}$, $2.0 \times 10^{19}$ and $3.0 \times 10^{19}\cm$.}

  \label{fig: he1048_mmap}
\end{figure}

As part of a program to bracket the distances to many HVCs, we have
obtained echelle spectra of two stars: \tee\ (which is aligned with WW\,35)
and \hee\ (WW\,62). Both stars were drawn from our catalog of Field
Horizontal Branch stars aligned with HVC gas
\citep{thom-etal-05-probes-catalogue}; Figure~\ref{fig: he1048_mmap} shows
their alignment with the HVC gas.  Basic information for both lines of
sight is listed in Table~\ref{tab: basic-info}.  Results for other targets
in this program will be reported in a future paper.

The stars \tee\ and \hee\ were observed in excellent conditions on 2005
April 01 and 2005 March 31 respectively, using the MIKE spectrograph
\citep{bernstein-etal-03-MIKE} mounted on the Magellan Clay 6.5m
telescope. Spectra covering the range 3350 -- 9500\,\AA\ with resolving
power (as measured from arc lines) $\mbox{R} = 40,000$ at 4000\,\AA\ and
$\mbox{R} = 31,000$ at 5900\,\AA\ were obtained.  The data were reduced
using the MIKE {\it redux} package\footnote{See http://web.mit.edu/\~{
}burles/www/MIKE/} and corrected from a heliocentric to Local Standard of
Rest (LSR) frame for comparison with \HI\ emission data.

\HI\ emission data were obtained from two sources. First we used the \HI\
spectra presented in \citet{thom-etal-05-probes-catalogue}, which was taken
from the \HI\ Parkes All-Sky Survey.  These data were reduced in such a way
as to be sensitive to extended emission \citep{putman-etal-03-mshvc}.
Secondly, to estimate the column density along the line of sight, we
obtained higher velocity resolution (but poorer spatial resolution) data
from the Leiden/Argentine/Bonn (LAB) survey \citep{kalberla-etal-05-LAB}.

Accurate \ubv\ photometry was obtained with the WIYN 0.9\,m telescope
between 2005 November 20 -- 22, with Landolt standards obtained at the same
time. The data were reduced in the usual fashion using {\it IRAF} and
aperture photometry was performed. The results are listed in
Table~\ref{tab: basic-info}.

\section{Results}
\label{sec: results}
\subsection{\rm{HE1048$+$0231}}
\label{sec: he1048}

\begin{figure*}
  \epsscale{1.0}
  \plottwo{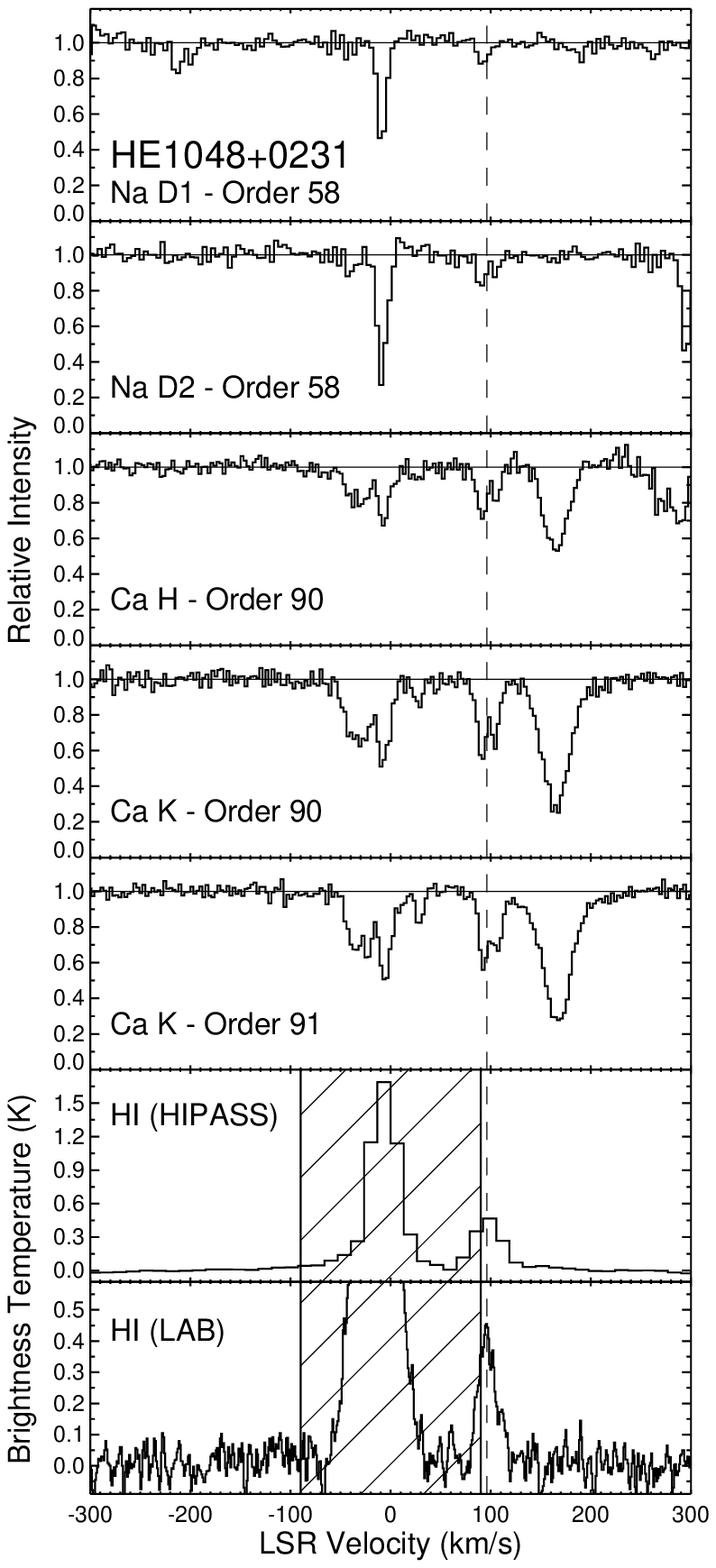}{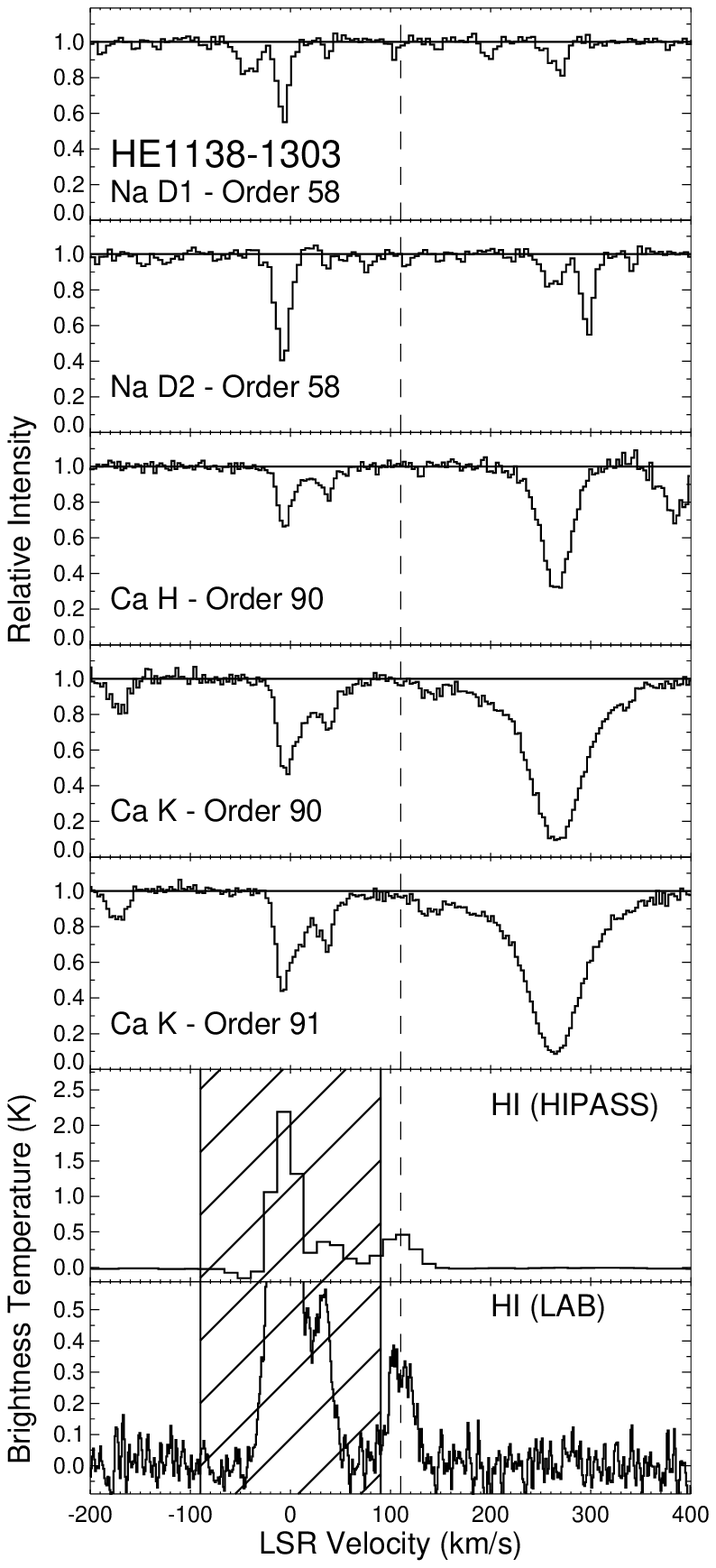}

  \caption{Optical and radio spectra along the line of sight towards \tee\
  (left panel) and \hee\ (right panel). For \tee, both \Ca\ lines, and at
  least one \NaI\ line, show the same absorption structure. The large
  absorption line at $\sim 170\kms$ is the intrinsic stellar \Ca\ line,
  while the complex absorption near 0\kms\ is due to local gas. For \hee,
  no such HVC absorption is present; the absorption line at $\sim 140\kms$
  near \CaK\ is an intrinsic \ion{Ti}{2} line. The shaded region in the
  \HI\ spectra show the limit of low velocity gas; below this the HIPASS
  data are expected to be unreliable.  The dashed vertical line is drawn at
  the velocity of the HVC, as measured in the LAB spectrum (see
  \S~\ref{sec: he1048}). }

  \label{fig: teehee_stack}
\end{figure*}

In the spectrum of \tee\ we detect a clear signature of absorption, shown
in Figure~\ref{fig: teehee_stack} (left panel). We plot the optical
absorption and \HI\ emission spectra aligned in velocity space. Both lines
of the \NaI\ doublet, the \CaHK\ lines, and the HIPASS and LAB \HI\ spectra
are shown. Note that two separate \CaK\ spectra are shown, since this line
falls onto two adjacent echelle orders, allowing us to confirm in one order
what is seen in the other. Significant HVC absorption ($> 3\sigma$) was
detected in all of these lines. Since the same absorption structure is seen
at the same velocity of several different species, we can reject the
possibility that these are intrinsic stellar features. The shaded region
indicates the boundary of lower velocities ($|\vLSR| \leq 90\kms$) where
HIPASS data are unreliable, while the vertical line shows the HVC velocity
($|\vLSR| = 96\kms$).


\begin{deluxetable}{lcc}
\tabletypesize{\scriptsize}
\tablecaption{\label{tab: basic-info}Basic parameters of HVC probes.
}
\tablewidth{8cm}
\tablecolumns{3}
\tablehead{
\colhead{} &
\colhead{\tee} &
\colhead{\hee}
}
\startdata
\sidehead{{\it Stellar Properties}}
RA,DEC (J2000)                           &  10:50:52.5 +02:15:19      & 11:41:14.8 $-$13:20:17       \\
($l$,$b$)                                &  (248.517, +51.911)        & (277.905, +46.098)           \\
Magnitude (V$_0$, (\bv)$_0$)             &  $16.51\pm0.03$; 0.00      & $14.87\pm0.01$; 0.40         \\
Distance (kpc)                           &  $8.8^{+2.3}_{-1.3}$       & $7.7 \pm 0.2$                \\
Exp time (s)                             &  $10 \times 1800$          & $3 \times 1800$              \\
Signal-to-noise (pix$^{-1}$)             &  60                        & 90                           \\    
\hline                                  
\sidehead{{\it HVC properties}}     
N(\HI) ($\times 10^{19}$ \cm)            & $1.64\pm0.31$              & $ 1.73\pm0.33$               \\
\vLSR (\kms)                             & 96                         & 110                          \\
W$_\lambda$ (\CaK)(\mA)                  & $116.2\pm 4.6  $           & $\sigma(\mbox{W}_{\lambda}) = 3.4$  \\
W$_\lambda$ (\CaK)(\mA)                  & $112.9\pm 4.1  $           & $\sigma(\mbox{W}_{\lambda}) = 3.9$  \\
W$_\lambda$ (\CaH)(\mA)                  & $ 69.5\pm 7.2  $           & $\sigma(\mbox{W}_{\lambda}) = 3.3$  \\
W$_\lambda$ (\NaD)(\mA)                  & $ 31.0\pm 7.0  $           & $\sigma(\mbox{W}_{\lambda}) = 4.4$  \\
W$_\lambda$ (\NaDb)(\mA)                 & $ 51.3\pm 7.7  $           & $\sigma(\mbox{W}_{\lambda}) = 5.0$  \\
\enddata

\tablecomments{Magnitudes have been corrected for dust extinction using the
maps of \citet{schlegel-etal-98-dust}. Distance errors are computed based
on the variation of atmospheric parameters (obtained from photometric
errors).}

\end{deluxetable}

Equivalent widths (EWs) and errors were derived following the method
outlined in \citet[][ hereafter SS92]{sembach-savage-92-EW}.  Legendre
polynomials (order 2 $-$ 5) were fit to the local continuum regions
($|\vLSR| < 1000\kms$), with an F-test determining the appropriate
polynomial order. EWs were measured by direct integration of the normalized
spectra in the range $78 - 115\kms$, as determined from the \CaK\
absorption. EW errors due to both intensity variations and continuum
placement were added in quadrature. The results are listed in
Table~\ref{tab: basic-info}. Note that the EWs of the \CaK\ and \CaH\ are
in the expected $\sim 2:1$ ratio (at the limit of the range permitted by
the errors), as is the EW ratio of \NaDb\ to \NaDa.


In order to disentangle the complex absorption structure, multiple
Gaussians were fit to the HVC absorption features.  Averaging over the 5
separate spectral regions, we determined line centers of 91.5 and
105.2\kms. A multi-component Gaussian decomposition was attempted for the
HVC emission seen in the LAB spectrum, but the data did not support more
than a single component, centered at $96\kms$.  We conclude that beam
smearing is a significant problem -- the \HI\ data have a spatial
resolution of 40 arcmin \citep{kalberla-etal-05-LAB}.  The HIPASS data,
while having higher spatial resolution, have a much coarser velocity
sampling, and were no help in this regard. This multiple absorption is
indicative of either two distinct, spatially aligned clouds, or two
components of the same cloud with a large kinematic separation. Higher
resolution data are required to fully disentangle the
possibilities.

Averaging over the two \CaK\ measurements, we obtained \teeNCa, assuming a
linear relation between equivalent width and column density
\citep[e.g.][]{savage-sembach-96-review}.  The corresponding total \HI\
column density (from the LAB data) is \teeNHI, where we have again assumed
the optically thin case \citep[since the \HI\ brightness temperature is
low, this assumption is valid;][]{dickey-lockman-90-MW-HI-review}.
Combining these two values yields a \Ca\ to \HI\ ratio of \teeNCaHI. The
solar \Ca\ abundance is A$_{\odot} = 2.2\times10^{-6}$
\citep{anders-grevesse-89-abundances}, yielding an abundance of $0.04$
solar.  Note that ionization effects (which are difficult to estimate) have
not been taken into account.

We attempted to confirm the low surface gravity nature of \tee\ using the
method of \citet{wilhelm-etal-99a-classification}. We derived stellar
parameters in the range $T_{\rm eff} = 9350 - 9750$\,K, $\mbox{log}\,{\it
g} = 3.5 - 4.25$, with an associated distance $\mbox{d} =
8.8^{+2.3}_{-1.3}\kpc$. Note that these values place \tee\ on the
main-sequence, with a color consistent with spectral type A0. The
uncertainty in $\mbox{log}\,{\it g}$ (and hence the distance) is due mainly
to the degeneracy of \ub\ in this temperature range.

\subsection{\rm{HE1138$-$1303}}
\label{sec: he1138}

The star \hee\ is aligned with cloud WW\,62. While not originally
classified as part of \wb, its position and velocity strongly suggest an
association. Note also that in this region, the complexes WA and WB inhabit
similar parts of position and velocity space (c.f. Figs 1j and 1k of
WvW91).  The interpretation of the optical spectrum of \hee\ (Fig~\ref{fig:
teehee_stack}, right panel) is less certain than that of \tee.  The same
methods as described above and in SS92 were applied, with the results given
in Table~\ref{tab: basic-info}. Integration was performed in the range $92
- 127\kms$ ($\pm 1.5$ the Gaussian $\sigma$ determined from the LAB
spectrum). No significant absorption near \CaH\ or K was detected.  There
is a single-pixel feature at the location of the \NaDa line which yields an
EW of \heeWNaDa. Around \NaDb\ there is an unidentified absorption feature
with \heeWNaDb. Neither of these are statistically significant detections,
nor are their locations correlated. Together with the lack of \CaK\
absorption, we conclude that there is no evidence of HVC absorption towards
\hee.

The interpretation of the lack of observed HVC absorption in the direction
of HE 1138-1303 requires some caution \citep[see comments
by][]{wakker-01-distance-metallicity}.  We first assumed that beam-smearing
would contribute a factor of two reduction in the \HI\ column density along
the sight line.  Next, metallicities are known to vary within a cloud
\citep{gibson-etal-01-ComplexC}. We thus assumed N(\Ca)/N(\HI) is a factor
of two lower than that measured in the \tee\ line of sight \cite[which is
unlikely when the PKS\,0837--12 sight line is considered;
see][]{robertson-etal-91-complexWB} . We allow a further factor of two
reduction to allow for ionization uncertainty.  Taking this into account,
we would then expect to observe a \CaK\ EW of $\sim15\mA$.  The measured
error in these orders is $\sim3.4\mA$, giving a ratio of
$\mbox{W}_{\lambda, expected}/\sigma(W_{\lambda}) \sim 4.4$, i.e. we expect
to see HVC absorption with at least 4$\sigma$ confidence (or more, since it
is unlikely that all the above factors would be perfectly correlated).

We also performed a stellar parameter analysis for \hee. The results
suggested stellar parameters $T_{\rm eff} = 6000$\,K, $\mbox{log}\,{\it g}
= 2.0$ (i.e.\ on the Horizontal Branch), at a distance of $\sim 7.7 \pm
0.2\kpc$. These parameters are near the limits of the models, so the errors
are likely to be somewhat underestimated.

\section{Discussion}
\label{sec: discussion}

The upper distance limit we obtain places WB firmly inside the halo of the
Milky Way. The non-detection towards \hee\ sets a probable lower limit.
Although a scenario in which the two clouds are physically disjoint is not
excluded, it is unlikely given the location and kinematics of the gas. A
distance of $7.7\kpc < \mbox{d} < 8.8\kpc$ implies a total \HI\ mass of
$3.8 \times 10^5\Msun < \mbox{M}_{\HI} < 4.9 \times 10^5\Msun$ for the
entire complex. We measured a \Ca\ abundance $N(\Ca)/N(\HI) =
81\pm16\times10^{-9}\cm$ ($\sim 0.04$ solar) which is about a factor of
four less than that seen towards PKS\,0837--120 \citep[$280 \times
10^{-9}$; ][]{wakker-01-distance-metallicity}, $\sim 35\degr$ away from
\tee.

Since the \HI\ data have poor spatial resolution (in the case of the LAB)
and poor velocity resolution (HIPASS), we cannot resolve individual cloud
structures; the \HI\ column density estimates are likely to be accurate
only to a factor of two.  While the broad characteristics of the \HI\ and
optical data match, we cannot resolve any condensations within the cloud,
which are suggested by the multiple absorption components seen in the
optical data. Higher resolution interferometric or large single-dish data
(e.g. ALFA at Arecibo), with good velocity resolution, are thus crucial.

In the Galactic rest frame, the measured velocity of \wb\ (96\kms) implies
that the complex is in-falling with $v_{\mbox{\scriptsize GSR}} = -30\kms$.
A maximum distance of 9\kpc\ places \wb\ $7\kpc$ above the disk of the
Milky Way, some $12\kpc$ from the Galactic center. Both fountain and
infalling gas interpretations are consistent with the data.  Under an
infall scenario, the anomalously high N(\Ca)/N(\HI) measurement might be
explained in terms of low depletion due to low dust content or hydrogen
ionization (which has not been taken into account here).  If the gas is
closer to the disk, then \wb\ may be fountain gas returning to the
disk. The \Ca\ abundance is not implausible since \Ca\ does not accurately
trace the total gas metallicity, although a more accurate metallicity
estimate would be very helpful for this discussion.

The distance bracket towards \wb\ is an important step towards the final
goal of determining the distances to a large sample of High Velocity
Clouds. It is one of only a few upper distance limits to HVCs, and the
first direct limit towards a cloud that was predicted to be very distant by
the \citet{blitz-etal-99} model.  Our results add to the growing body of
evidence that the High Velocity Clouds are much more likely to belong to a
Galactic or circum-Galactic population, rather than a distant,
extra-Galactic one.

\acknowledgments

We are grateful to Rebecca Bernstein, Ian Thompson and Steve Shectman for
help with the MIKE spectrograph. We thank Scott Burles and Jason Prochaska
for help with the MIKE redux package. Thanks to Peter Kalberla for
providing the LAB survey data before publication. Nathan DeLee participated
in the WIYN observations, for which we are grateful. CT would like to thank
Tuorla Observatory for its support, and acknowledges travel support from
the ANSTO AMRF Programme. The financial support of the Australian Research
Council is greatly appreciated. TCB acknowledges partial support NSF grants
AST 04-06784 and PHY 02-16783.


\bibliography{AbsorptionLines,Abundances,BHB,books,CHVC,Dust,FUSE,GalacticFountain,Halo,HES,HI,hipass,HVC,Instrumentation,ISM,LocalGroup,mary,me,MetalPoor,ModelAtmospheres,nc,wakker,CDM,HVC-Halpha,theses,ChemicalEvolution}

\begin{thebibliography}{25}
\expandafter\ifx\csname natexlab\endcsname\relax\def\natexlab#1{#1}\fi

\bibitem[{{Anders} \& {Grevesse}(1989)}]{anders-grevesse-89-abundances}
{Anders}, E. \& {Grevesse}, N. 1989, \gca, 53, 197

\bibitem[{{Bates} {et~al.}(1990){Bates}, {Catney}, \&
  {Keenan}}]{bates-etal-90-HVC-dist}
{Bates}, B., {Catney}, M.~G., \& {Keenan}, F.~P. 1990, \mnras, 242, 267

\bibitem[{{Bernstein} {et~al.}(2003){Bernstein}, {Shectman}, {Gunnels},
  {Mochnacki}, \& {Athey}}]{bernstein-etal-03-MIKE}
{Bernstein}, R., {Shectman}, S.~A., {Gunnels}, S.~M., {Mochnacki}, S., \&
  {Athey}, A.~E. 2003, in Proc. SPIE, Vol. 4841, 1694

\bibitem[{{Blitz} {et~al.}(1999){Blitz}, {Spergel}, {Teuben}, {Hartmann}, \&
  {Burton}}]{blitz-etal-99}
{Blitz}, L., {Spergel}, D.~N., {Teuben}, P.~J., {Hartmann}, D., \& {Burton},
  W.~B. 1999, \apj, 514, 818

\bibitem[{{Bregman}(1980)}]{bregman-80}
{Bregman}, J.~N. 1980, \apj, 236, 577

\bibitem[{{Danly} {et~al.}(1993){Danly}, {Albert}, \&
  {Kuntz}}]{danly-etal-93-ComplexM}
{Danly}, L., {Albert}, C.~E., \& {Kuntz}, K.~D. 1993, \apjl, 416, L29+

\bibitem[{{Dickey} \& {Lockman}(1990)}]{dickey-lockman-90-MW-HI-review}
{Dickey}, J.~M. \& {Lockman}, F.~J. 1990, \araa, 28, 215

\bibitem[{{Gibson} {et~al.}(2001){Gibson}, {Giroux}, {Penton}, {Stocke},
  {Shull}, \& {Tumlinson}}]{gibson-etal-01-ComplexC}
{Gibson}, B.~K., {Giroux}, M.~L., {Penton}, S.~V., {Stocke}, J.~T., {Shull},
  J.~M., \& {Tumlinson}, J. 2001, \aj, 122, 3280

\bibitem[{{Kalberla} {et~al.}(2005){Kalberla}, {Burton}, {Hartmann}, {Arnal},
  {Bajaja}, {Morras}, \& {P{\"o}ppel}}]{kalberla-etal-05-LAB}
{Kalberla}, P.~M.~W., {Burton}, W.~B., {Hartmann}, D., {Arnal}, E.~M.,
  {Bajaja}, E., {Morras}, R., \& {P{\"o}ppel}, W.~G.~L. 2005, \aap, 440, 775

\bibitem[{{Klypin} {et~al.}(1999){Klypin}, {Kravtsov}, {Valenzuela}, \&
  {Prada}}]{klypin-etal-99-missing-satellites}
{Klypin}, A., {Kravtsov}, A.~V., {Valenzuela}, O., \& {Prada}, F. 1999, \apj,
  522, 82

\bibitem[{{Oort}(1966)}]{oort-66-HVC-origins}
{Oort}, J.~H. 1966, \bain, 18, 421

\bibitem[{{Putman} {et~al.}(2003){Putman}, {Staveley-Smith}, {Freeman},
  {Gibson}, \& {Barnes}}]{putman-etal-03-mshvc}
{Putman}, M.~E., {Staveley-Smith}, L., {Freeman}, K.~C., {Gibson}, B.~K., \&
  {Barnes}, D.~G. 2003, \apj, 586, 170

\bibitem[{{Robertson} {et~al.}(1991){Robertson}, {Schwarz}, {van Woerden},
  {Murray}, {Morton}, \& {Hulsbosch}}]{robertson-etal-91-complexWB}
{Robertson}, J.~G., {Schwarz}, U.~J., {van Woerden}, H., {Murray}, J.~D.,
  {Morton}, D.~C., \& {Hulsbosch}, A.~N.~M. 1991, \mnras, 248, 508

\bibitem[{{Savage} \& {Sembach}(1996)}]{savage-sembach-96-review}
{Savage}, B.~D. \& {Sembach}, K.~R. 1996, \araa, 34, 279

\bibitem[{{Schlegel} {et~al.}(1998){Schlegel}, {Finkbeiner}, \&
  {Davis}}]{schlegel-etal-98-dust}
{Schlegel}, D.~J., {Finkbeiner}, D.~P., \& {Davis}, M. 1998, \apj, 500, 525

\bibitem[{{Sembach} \& {Savage}(1992)}]{sembach-savage-92-EW}
{Sembach}, K.~R. \& {Savage}, B.~D. 1992, \apjs, 83, 147

\bibitem[{{Sembach} {et~al.}(1991){Sembach}, {Savage}, \&
  {Massa}}]{sembach-savage-massa-91-HVC-dist}
{Sembach}, K.~R., {Savage}, B.~D., \& {Massa}, D. 1991, \apj, 372, 81

\bibitem[{{Shapiro} \& {Field}(1976)}]{shapiro-field-76}
{Shapiro}, P.~R. \& {Field}, G.~B. 1976, \apj, 205, 762

\bibitem[{{Thom} {et~al.}(2005){Thom}, {Gibson}, \&
  {Christlieb}}]{thom-etal-05-probes-catalogue}
{Thom}, C., {Gibson}, B.~K., \& {Christlieb}, N. 2005, \apjs, 161, 147

\bibitem[{{van Woerden} {et~al.}(1999){van Woerden}, {Schwarz}, {Peletier},
  {Wakker}, \& {Kalberla}}]{vanwoerden-etal-99-nature}
{van Woerden}, H., {Schwarz}, U.~J., {Peletier}, R.~F., {Wakker}, B.~P., \&
  {Kalberla}, P.~M.~W. 1999, \nat, 400, 138

\bibitem[{{Wakker}(1991)}]{wakker-91-II-distribution}
{Wakker}, B.~P. 1991, \aap, 250, 499

\bibitem[{{Wakker}(2001)}]{wakker-01-distance-metallicity}
---. 2001, \apjs, 136, 463

\bibitem[{{Wakker} \& {van Woerden}(1991)}]{wvw91}
{Wakker}, B.~P. \& {van Woerden}, H. 1991, \aap, 250, 509

\bibitem[{{Wannier} {et~al.}(1972){Wannier}, {Wrixon}, \&
  {Wilson}}]{wannier-wrixon-wilson-72}
{Wannier}, P., {Wrixon}, G.~T., \& {Wilson}, R.~W. 1972, \aap, 18, 224

\bibitem[{{Wilhelm} {et~al.}(1999){Wilhelm}, {Beers}, \&
  {Gray}}]{wilhelm-etal-99a-classification}
{Wilhelm}, R., {Beers}, T.~C., \& {Gray}, R.~O. 1999, \aj, 117, 2308

\end{thebibliography}
\bibliographystyle{apj}

\end{document}